\documentclass[vecphys]{svmult}

\usepackage{graphicx}        
\usepackage[bottom]{footmisc}

\begin{document}

\title*{Revealing the Nature of Asymmetric Planetary Nebulae through 
Abundance Analysis
}

\author{Letizia Stanghellini}
\institute{National Optical Astronomy Observatory,
950 N. Cherry Avenue, Tucson AZ 
857129; lstanghellini@noao.edu}

\titlerunning{PNe Abundances and Evolutionary Yields}

\maketitle

\begin{abstract}
The correlations between planetary nebula (PN) morphology and the nature of their progenitors
are explored by examining homogeneous PN samples in the Galaxy and the Magellanic Clouds.
We selected PNe with reliable abundances from spectral analysis, and whose morphology is known, 
and compared the abundances of the element at variance with stellar evolution with the final yields
of Asymptotic Giant Branch (AGB) stellar models. We found that most asymmetric PNe derive from the evolution of
massive AGB stars both in the Galactic disk and the Magellanic Clouds.

\end{abstract}

\section{Introduction}
Planetary Nebulae (PNe) are important probes of stellar evolution. The Asymptotic Giant Branch (AGB)
stellar envelopes carry the signature of the elemental evolution within the star, which in turn
depends on its initial metallicity and, predominantly, its main sequence mass. 
By comparing PN abundances of the elements at variance with AGB evolution, such as helium, carbon, oxygen, and nitrogen,
to evolutionary yields calculated for the final envelope ejection in stars of appropriate metallicity, 
one can set strong constraints to the initial stellar mass. The only assumption is that the PN progenitors
are single stars, or members of wide binary systems. Close binary stellar evolution is discussed elsewhere 
(see Izzard, this conference).

In this paper we compare abundances of Galactic and Magellanic Cloud PNe to the yields derived from single
star evolution. To this end, we selected PN with known morphology, so that we can follow systematic relations between 
the possible AGB origin of PN of different morphological classes.

\section{Selected Samples}

We use the Galactic disk PNe as in Stanghellini et al. [1], consisting of a large PN sample
whose
abundances were homogeneously recalculated from selected published
flux data. We also used the sample by [2] for carbon abundances, not available in [1]. 
Other authors have published samples of Galactic PN abundances. In the cases where the calculations are based on 
similar ionization correction factors [3], the elemental abundances compare well
with those used here [1]. The additional value of the sample used is that it includes only disk PNe,
explicitly excluding the halo and bulge populations.
Galactic PN morphological classes
are selected from the IAC Morphological Catalog of Northern Galactic PNe [4]. 

The LMC and SMC PN abundances used in this paper are from references
[5], [6], [7], [8], 
[9], and [10]. Morphology of Magellanic Cloud PNe, available only 
through {\it Hubble Space Telescope} observations, are from [11], [12], [13], and [14].

\begin{table}
\begin{tabular}{llrrr}
\hline\noalign{\smallskip}
  
{Element~~~~~~~~~~}   &   PN type& {~~~~~MW Disk} & {~~~~~LMC}& {~~~~~SMC}  \\

\noalign{\smallskip}\hline\noalign{\smallskip}

He/H&	Whole sample&		0.12&	0.10&	0.091\\ 
&Symmetric& 	0.11& 	0.09& 	0.086\\
&Asymmetric& 	0.15&	0.10&	0.092\\

&&&&\\
&&&&\\
C/H ($\times 10^4$)&	Whole sample&		5.7&	3.3&	2.8\\
&Symmetric& 	$\dots$& 5.2&	3.3\\
&Asymmetric&	$\dots$& 2.0&	$\dots$\\
&&&&\\
&&&&\\

N/H ($\times 10^4$)&	Whole sample&	2.4&	0.97&	0.46\\
&Symmetric&	1.6&	0.67&	0.30\\
&Asymmetric&	4.6&	1.5&	0.65\\
&&&&\\
&&&&\\
O/H ($\times10^4$)&	Whole sample&	3.5&	1.9&	1.1\\
&Symmetric&	3.4&	2.0&	1.5\\
&Asymmetric&	3.8&	1.0&	0.8\\
&&&&\\
&&&&\\
N/O&  Whole sample&		0.66&	0.62&	0.6\\
&Symmetric&	0.42&	0.36&	0.16\\
&Asymmetric&	1.32&	0.93&	0.91\\

\noalign{\smallskip}\hline
\end{tabular}
\caption{Average abundances of morphologically selected PNe in the Galactic disk, the LMC, and the SMC}
\end{table}

\section{Galactic Planetary Nebulae}

In Table 1, column (3), we show the average He/H, C/H, N/H, O/H, and N/O abundances of Galactic disk PNe.
All abundances are expressed linearly, the C/H, N/H, and O/H have been multiplied by 10$^4$. 
We used the Manchado et al. [15] classification scheme to group the PNe as symmetric (round and elliptical)
and 
asymmetric (bipolar, bipolar core, and quadrupolar).
We calculate the average abundances for each population and, where possible, 
for the symmetric and asymmetric PN classes within each population as well.
From the Table we infer that He/H and N/H averages are always higher in asymmetric than
symmetric PNe, both the Galactic disk
and the Magellanic Clouds. 

\begin{figure}[htp]
\centering\includegraphics[width=0.49\textwidth,viewport=20 100 600 800,clip]{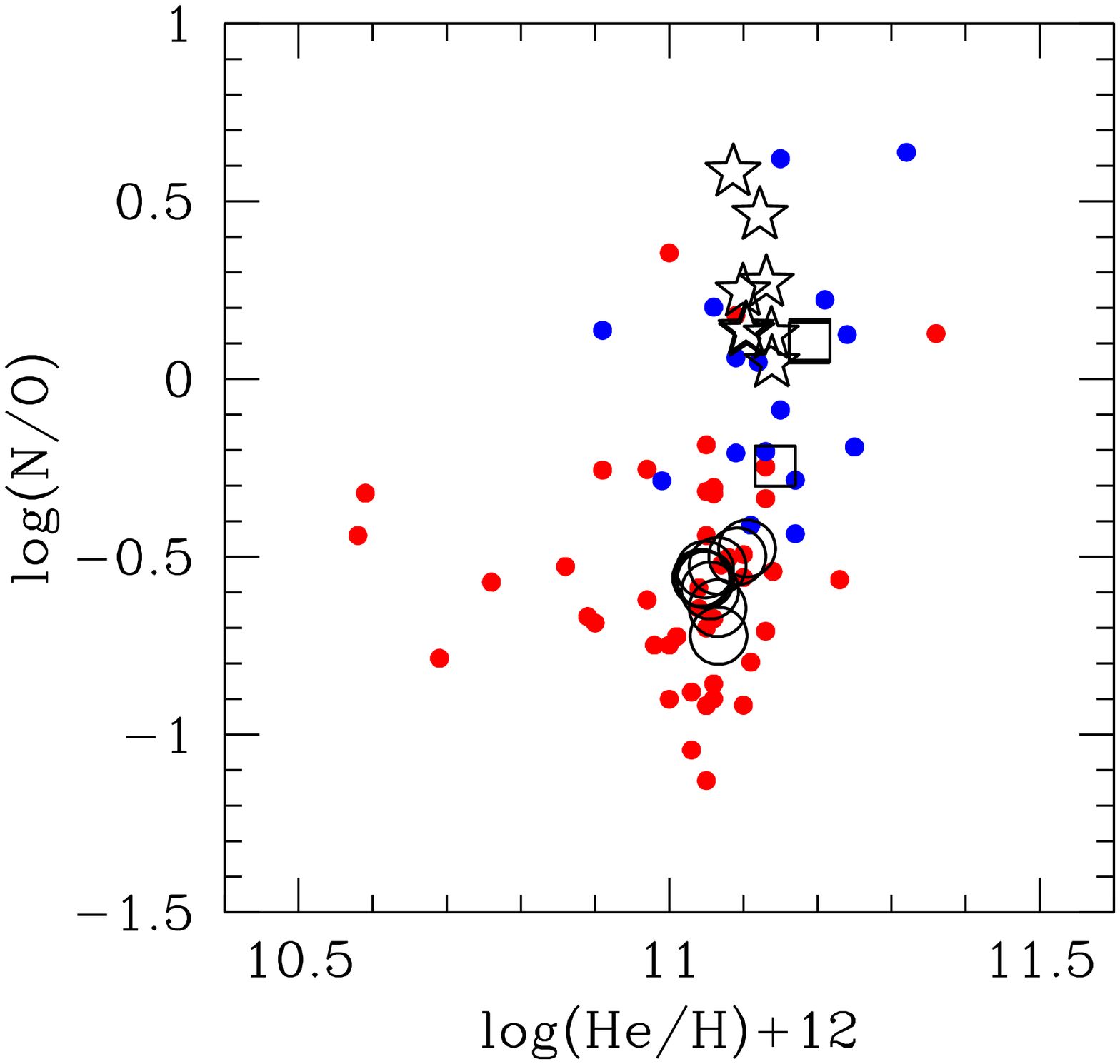}
\hfill
\includegraphics[width=0.49\textwidth,viewport=20 100 600 800,clip]{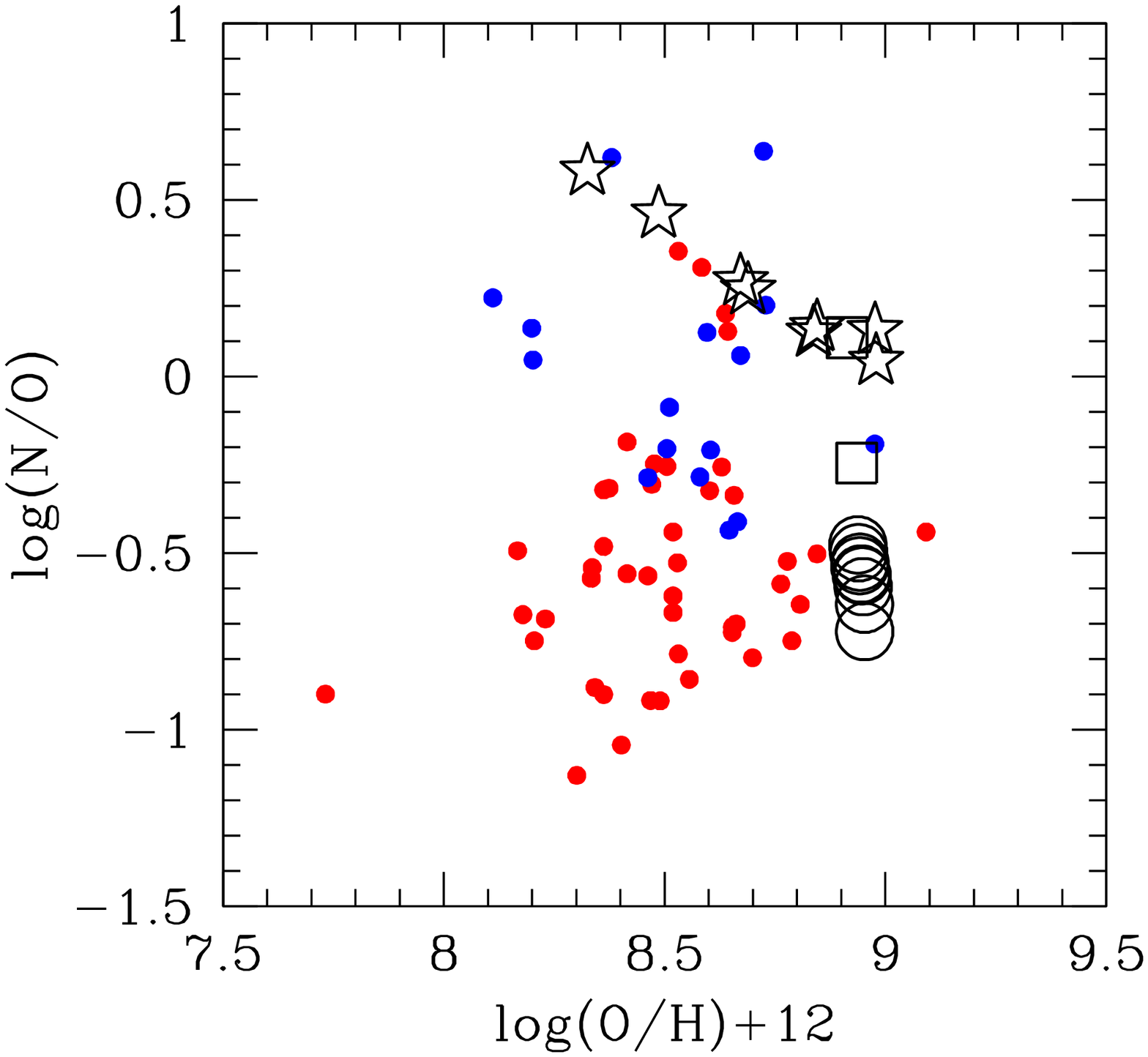}\\
\caption[]{Left panel: N/O vs. He/H; right panel: N/O vs. O/H. In both panels 
we plot the Galactic sample PNe, with Symmetric PNe represented by red dots, and Asymmetric PNe with blue dots. 
The overlapped open 
symbols represent the yields from stellar evolution: Open circles: 1$<$M/M$_{\odot}<$4 [16]; open squares
4$<$ M/M$_{\odot}<$6 [16]; open stars: 5$<$M/M$_{\odot}<$8 [17].}

\end{figure}

Figure 1 shows the comparison between the Galactic PN abundances and the yields
from stellar evolution. On the left panel we show the N/O vs. He/H abundances for the PNe in our sample.
Asymmetric PNe (blue dots) occupy a different locus than symmetric PNe (red dots), even if a small overlap of the two samples
exists. 
The yields resulting from  the envelope ejection of the evolution of low mass
(1$<$M/M$_{\odot}<$4) and intermediate mass (4$<$M/M$_{\odot}<$6) stars calculated by Karakas 
\& Lattanzio ([16], and Karakas, this volume) are indicated with open circles and squares respectively.
We interpret the plot in the sense that 
symmetric (i.e., round and elliptical) PNe originate from 
single star evolution with 
M$<$4 M$_{\odot}$, and most asymmetric PN are the progeny of evolution of M$>$4 M$_{\odot}$ stars.
The exception to this single (or wide binary) stellar scenario is represented by close binary evolution, which would
likely produce asymmetric PNe with low N/H (see Izzard, this volume).
We also plot 
the 5$<$M/M$_{\odot}<$8 yields from Gavilan et al. [17] with open stars. While Gavilan's results showing the
high mass yields well encompass the extremely high N/H asymmetric PNe,
these are based on synthetic rather than evolutionary models, thus should be 
be interpreted with caution.

In the right panel of Figure 1 we show N/O vs. O/H. The colors and symbols have the 
same meaning than in the left panel. Once more, the symmetric PNe seem to 
be well encompassed by low-mass star progenitors, and asymmetric PNe lie with the 
high mass yields of stellar evolution. 
It seems that the available models encompass only one range of initial metallicities, while 
Galactic PNe spread over a larger range.
The N/O vs. O/H plot gives and indication of the efficiency of the ON cycle in AGB stars. Following the models, it is clear
that the ON cycle is active for the high mass stars only, those going through the third dredge up.

\section{Magellanic Cloud PNe}

The same relations between the yields from stellar evolution and PN abundances have been found
in the Magellanic Cloud PNe, with the difference that the limiting mass for symmetric vs. asymmetric PN progenitors
is lower in the Clouds than in the Galactic disk. We infer that there is a tight correlation between high He/H and N/O ratios
and 
asymmetric morphology in all galaxies where it has been possible to test this relation. The comparison of the data with the
models of appropriate initial metallicity shows that the locus of asymmetric PNe in the diagnostic plots (such as those in Figure 1)
is encompassed by the yields for higher AGB mass stellar models. In Figure 2 we summarize our abundance data, by showing the
He/H and N/O ratios as averages for the different morphological classes, for Galactic and Magellanic Cloud PNe. 
The He/H and N/O averages are different in symmetric and asymmetric PNe, and their
ranges (indicated by the bars) barely overlap, in all three populations. The N/O ratio is much higher
in  asymmetric than symmetric PNe of all three populations. It is worth noting that the N/O ratio is unexpectedly high in the asymmetric
SMC PNe, likely an effect of oxygen depletion during the ON cycle that is very efficient at low metallicites.

\begin{figure}[htp]
\centering
\includegraphics[width=0.49\textwidth, viewport=20 100 600 800,clip]{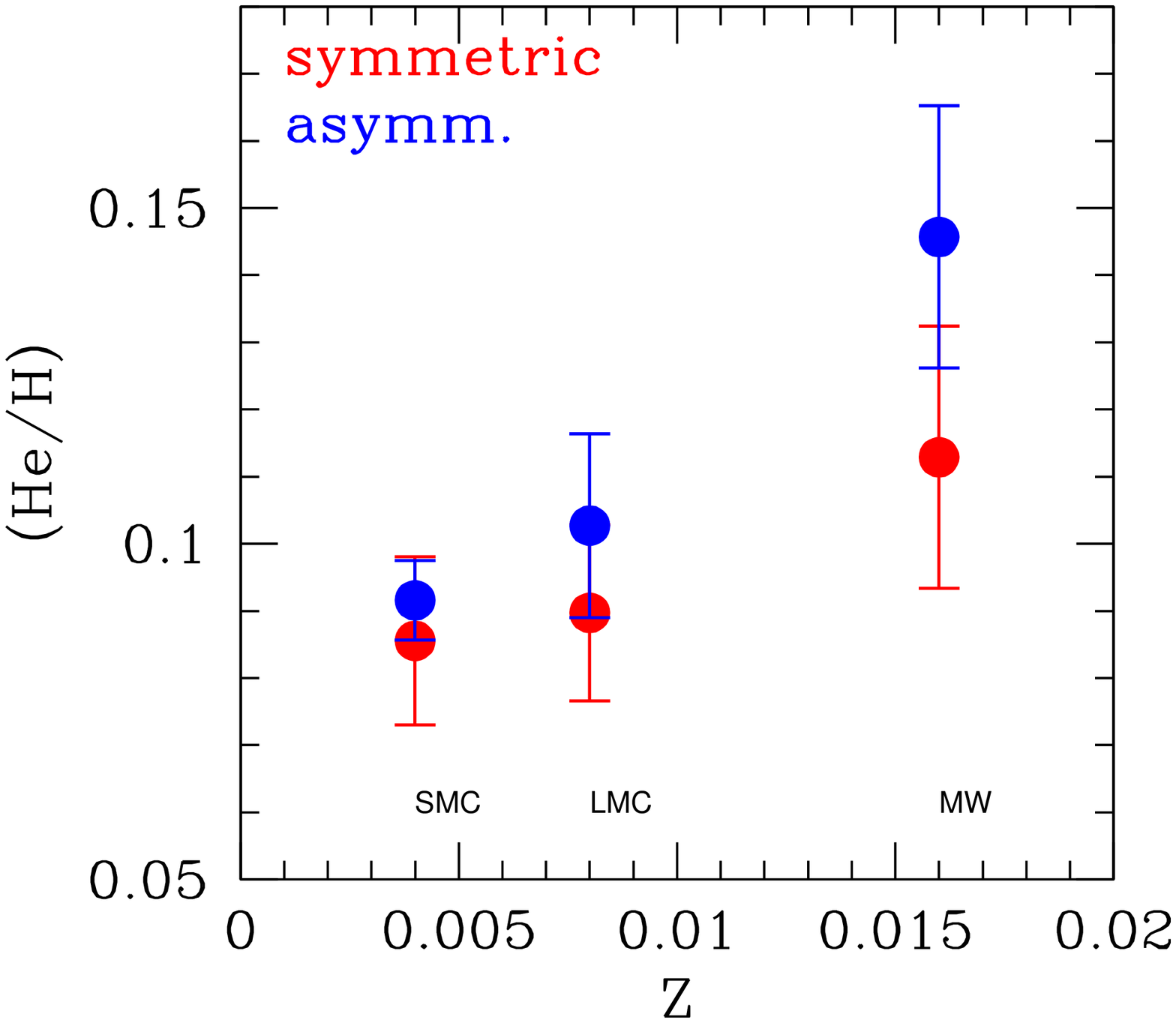}
\hfill
\includegraphics[width=0.49\textwidth,viewport=30 100 600 800,clip]{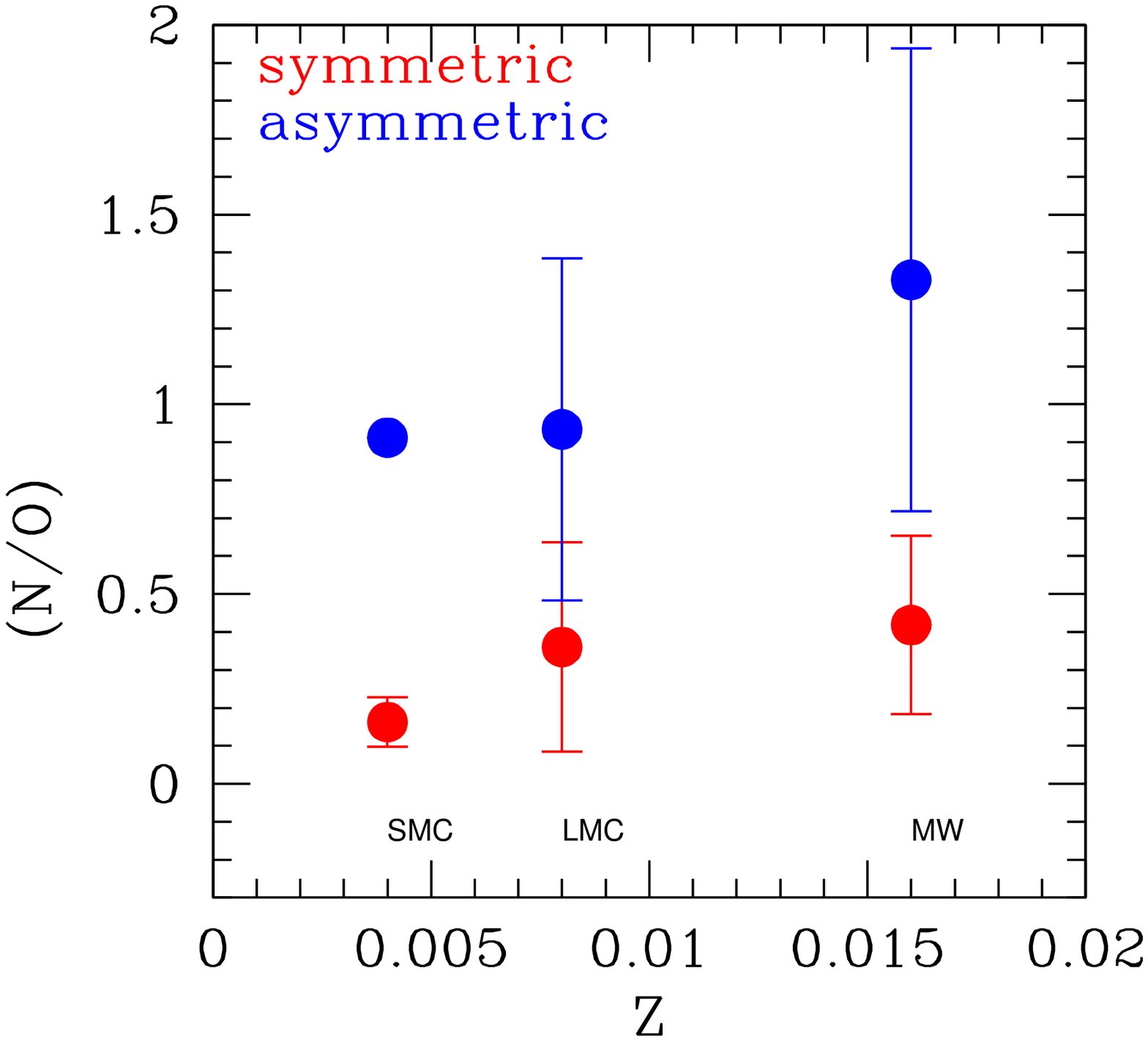}\\
\hfill
\caption[]{Averages (dots) and ranges (bars) of the He/H (left panel) and the N/O (right panel)
ratios in the three populations studied. The abscissa represents the average metallicity of the galaxy considered. 
We separate the symmetric (red) and asymmetric (blue) PN population.}
\end{figure}

\section{Final remarks}

With the straightforward selection of three PN populations from the Galactic disk, the LMC, and the SMC, and the assumption
that their elemental abundances derive from the stellar processing of material during the evolution of their progenitors, we
were able to infer that symmetric PNe mostly derive from the evolution of low mass ($<$4 M$_{\odot}$ in the Galaxy) AGB stars, 
while
the opposite is true for asymmetric PNe. The results
of this paper apply to the majority of the spatially resolved planetary nebulae, and exclude the products of close binary 
interactions.

\section{Acknowledgements}

I thank Bruce Balick, Katia Cunha, Martin Guerrero, Ting-Hui Lee,  Arturo Manchado, Richard Shaw, and Eva Villaver
for their contribution to the planetary nebula abundances and morphology projects. 
Many thanks to Amanda Karakas for providing her models in advance of publication, and 
to Amanda Karakas and Robert Izzard for scientific discussion.


\begin{thebibliography}{99.}


\bibitem[1]{2006ApJ...651..898S} Stanghellini, L., 
Guerrero, M.~A., Cunha, K., Manchado, A., \& Villaver, E.\ 2006, ApJ, 651, 
898 

\bibitem[2]{2004AJ....127.2284H} 
  Henry, R.B.C., Kwitter, K.\,B., \& Balick, B. ~2004, 
  AJ, 127, 2284 

\bibitem[3]{1994MNRAS.271..257K} Kingsburgh, 
R.~L., \& Barlow, M.~J.\ 1994, MNRAS, 271, 257 

\bibitem[4]{1996iacm.book.....M} Manchado, A., 
Guerrero, M.~A., Stanghellini, L., \& Serra-Ricart, M.\ 1996, The IAC 
morphological catalog of northern Galactic planetary nebulae, Publisher: La 
Laguna, Spain: Instituto de Astrofisica de Canarias (IAC), 1996, Foreword 
by Stuart R.~Pottasch, ISBN: 8492180609,  

\bibitem[5]{1996A&AS..116...95L} 
  Leisy, P. \& Dennefeld, M. ~1996, A\&AS, 116, 95 

\bibitem[6]{2005ApJ...622..294S} Stanghellini, L., 
Shaw, R.~A., \& Gilmore, D.\ 2005, ApJ, 622, 294 

\bibitem[7]{1989ApJ...339..872H} 
  Henry, R.B.C., Liebert, J., \& Boroson, T.\,A. ~1989, 
  ApJ, 339, 872 

\bibitem[8]{1988MNRAS.234..583M} 
  Monk, D.\,J., Barlow, M.\,J., \& Clegg, R.E.S. ~1988, 
 MNRAS, 234, 583 

\bibitem[9]{1989ApJ...339..844B} 
  Boroson, T.\,A. \& Liebert, J. ~1989, ApJ, 339, 844 

\bibitem[10]{1998A&A...336..667S} 
  Stasi\'nska, G., Richer, M.\,G., \& McCall, M.\,L. ~1998, 
  A\&A, 336, 667 

\bibitem[11)]{2001ApJ...548..727S} Shaw, R.~A., Stanghellini, 
L., Mutchler, M., Balick, B., \& Blades, J.~C.\ 2001, ApJ, 548, 727 

\bibitem[12]{2006ApJS..167..201S} Shaw, R.~A., Stanghellini, 
L., Villaver, E., \& Mutchler, M.\ 2006, ApJS, 167, 201 

\bibitem[13]{1999ApJ...510..687S} Stanghellini, L., 
Blades, J.~C., Osmer, S.~J., Barlow, M.~J., \& Liu, X.-W.\ 1999, ApJ, 510, 
687 

\bibitem[14]{2003ApJ...596..997S} Stanghellini, L., 
Shaw, R.~A., Balick, B., Mutchler, M., Blades, J.~C., \& Villaver, E.\ 
2003, ApJ, 596, 997 

\bibitem[15]{2000ASPC..199...17M} Manchado, A., 
Villaver, E., Stanghellini, L., \& Guerrero, M.~A.\ 2000, Asymmetrical 
Planetary Nebulae II: From Origins to Microstructures, 199, 17 


\bibitem[16]{2007arXiv0708.4385K} Karakas, A.~I., 
\& Lattanzio, J.~C.\ 2007, ArXiv e-prints, 708, arXiv:0708.4385 

\bibitem[17]{2006A&A...450..509G} Gavil{\'a}n, M., 
Moll{\'a}, M., \& Buell, J.~F.\ 2006, A\&A, 450, 509 





  







\end{thebibliography}
\end{document}